\documentclass[global,twocolumn]{svjour}
\usepackage{graphics}
\journalname{Applied Physics B}
\begin{document}
\title{Efficient fiber in-line single photon source based on colloidal single quantum dots on an optical nanofiber}
\author{K. Muhammed Shafi\inst{1} \and Kali P. Nayak\inst{1} \thanks{\emph{Corresponding author:} kalipnayak@uec.ac.jp} \and Akiharu Miyanaga\inst{2} \and Kohzo Hakuta\inst{1}
%
}                     
%
%
\institute{Center for Photonic Innovations and Institute for Laser Science, University of Electro-Communications, Chofu-Shi, Tokyo 182-8585, Japan. \and NS Materials Inc., Tsukushino, Fukuoka 818-0042, Japan.}
\date{Received: date / Revised version: date}
%
\maketitle
\begin{abstract}
We demonstrate a fiber in-line single photon source based on a hybrid system of colloidal single quantum dots deposited on an optical nanofiber and cooled down to cryogenic temperature (3.7 K). We show that a charged state (trion) of the single quantum dot exhibits a photo-stable emission of single photons with high quantum efficiency, narrow linewidth (3 meV FWHM) and fast decay time ($10.0\pm0.5$ ns). The single photons are efficiently coupled to the guided modes of the nanofiber and eventually to a single mode optical fiber. The brightness (efficiency) of the single photon source is estimated to be $16\pm2\%$ with a maximum photon count rate of $1.6\pm0.2$ MHz and a high single photon purity ($g^2(0)=0.11\pm0.02$). The device can be easily integrated to the fiber networks paving the way for potential applications in quantum networks. 
\end{abstract}
\section{Introduction}

Single photons play a central role in the realization of quantum information science \cite{kimble,OBrien}. Therefore, development of an ideal single photon source (SPS) is a long-standing frontier of research and currently a technology on high demand. There have been significant advances and innovations in this direction, albeit an ideal and on-demand SPS is yet to be realized \cite{Aharonovich,Senellart}. 

A straightforward approach to realize an SPS is to collect emission from an isolated quantum emitter. Neutral atoms or ions may be an orthodox choice for a quantum emitter. However, isolating single atoms or ions still requires a complex experimental apparatus. 
In this context, solid-state quantum emitters like quantum dots (QDs) or atom-like defect in crystalline hosts \cite{Aharonovich,Senellart,Michler} are one of the promising choices for practical SPS, based on the excellent emission properties and easier techniques to isolate single emitters. However, the presence of the complex mesoscopic solid-state host induces various challenges, e.g. inhomogeneous spectral broadening, emission intermittency and non-identical emitters, etc. In the last decade, there have been significant efforts and developments to overcome such challenges \cite{Aharonovich,Senellart,Park,Lin}.

On the other hand, a key challenge in this approach is to efficiently collect the single photons emitted from the individual emitter into a single spatial mode. Moreover, from the view point of applications in quantum networks, the single photons must be efficiently coupled to a single mode optical fiber (SMF). Advances in the nanophotonics and nanofabrication have shown promising developments in this regard. Excellent ideas and technologies have emerged to manipulate the fluorescence of single emitters using nanophotonic platforms like nano-waveguides, photonic crystal waveguides or cavities, plasmonic structures and other microresonators \cite{Aharonovich,Senellart,BramatiNanorod,BramatiNanorod2,Sipahigil,Daveau,Hoang,Waks}. Efficient collection of single emitter fluorescence into the nanophotonic platforms has been successfully demonstrated. However, coupling single photons from nanophotonic platforms to a SMF with high efficiency still remains a challenge \cite{Daveau,Waks}.    

In this context, tapered optical fiber with subwavelength diameter waist, optical nanofiber (ONF), provides a unique fiber in-line platform for collecting single emitter fluorescence \cite{famsan1,Kali1}. The distinct point of the technique is that along with the strong photonic confinement, the guided mode of the ONF can adiabatically evolve to the SMF mode with near unity efficiency. This provides an automatic and alignment-free fiber-coupled platform for an SPS. In the last decade, there has been significant development to interface the ONF platform with single emitters in the form of laser-cooled atoms \cite{Kali2,Kali3}, QDs \cite{Yalla2,FujiwaraQD}, color centers in nano-diamond \cite{ONFNDarno,Fujiwara}, 2D materials \cite{Schell2D} and molecules \cite{Skoff}. 

In particular, using colloidal single QDs on the ONF, it was demonstrated that the channeling efficiency can reach the theoretical limit of 22\% \cite{famsan1,Yalla1}, which is determined by the transverse confinement of the photonic mode in the ONF. Furthermore, it was demonstrated that the channeling efficiency can be enhanced to 65\% using a composite photonic crystal ONF cavity \cite{famsan2,Yalla3}. However, the experiments were performed in the room temperature resulting in a broadband emission from the QD. 

The key requirements for an efficient SPS are that the quantum emitter must be bright, producing high single photon count rate with high quantum efficiency (QE) and preferably in a narrow spectral window \cite{Aharonovich,Senellart}. Various quantum photonics protocols require narrow band (in ideal case, \textit{indistinguishable}) single photons \cite{Aharonovich,Senellart}. In this context, cryogenic operation of the ONF/QD hybrid system may suppress the phonon mediated broadening of the QD emission, leading to a narrow band SPS. 

We have reported the cryogenic cooling of the ONF/QD hybrid system to a temperature of 3.7 K \cite{Shafi1,Shafi2}. A gradient thick shell type CdSe QDs with an outermost ZnS shell, was used for the experiments. The spectroscopic studies at the cryogenic temperature had revealed that the photo-luminescence (PL) emission occurs not only from the neutral exciton but also from the charged exciton known as trion \cite{Shafi1,Shafi2,Louyer}. Moreover, from the PL decay profile, it was inferred that the upper state of the neutral exciton, consists of two levels with a dark state. This leads to an effectively slow decay component for the neutral exciton \cite{Shafi1,Biadala}. While, the trion state behaves effectively as a two-level system and the decay rate is one order faster than the effective slow decay component of the neutral exciton \cite{Shafi1,Fernee1,Javaux}. Therefore, the presence of the dark state of the neutral exciton hinders the implementation of the system as an efficient SPS. 

Recently, we have found that at cryogenic temperature, the CdSe QDs can be photo-charged by irradiating with shorter wavelength lasers. In particular, the QDs can be prepared in a permanently charged state using a 355 nm laser which eradicates the issues related to the dark state of the neutral exciton. A systematic investigation of the photo-charging behavior of the QDs measured by varying the wavelength and the intensity of the irradiating laser will be reported elsewhere. 

Here, we demonstrate the capability of the ONF/QD hybrid system as an efficient fiber in-line SPS. We show that at cryogenic temperature (3.7 K), the charged single QD exhibits a photo-stable emission of single photons with high QE, narrow linewidth (3 meV FWHM) and fast decay time ($10.0\pm0.5$ ns). The single photons are efficiently coupled to the guided modes of the ONF and eventually to a SMF. The brightness (efficiency) of the SPS is estimated to be 16$\pm2\%$ with a maximum photon count rate of 1.6$\pm0.2$ MHz and a high single photon purity ($g^2(0)=0.11\pm0.02$).

\begin{figure}[t!]
\centering
\resizebox{0.5\textwidth}{!}{%
  \includegraphics{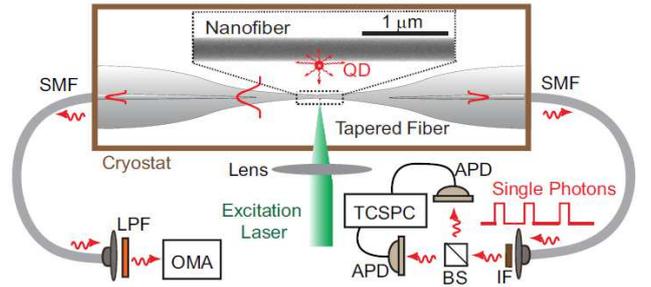}
}
\caption{Schematic diagram of the fiber in-line single photon source. The ONF/QD hybrid system is cooled down in a cryostat. The single QD is irradiated with a pulsed laser launched through the optical window of the cryostat. The emission from the single QD is coupled to the guided mode of the ONF and eventually to a single mode optical fiber (SMF). The temporal and spectral characteristics of the fiber-guided photons are measured simultaneously at the opposite ends of the fiber using a time correlated single photon counter (TCSPC) and an optical multi-channel spectrum analyzer (OMA), respectively. The abbreviations LPF, IF, BS and APD denote long-pass filter, interference filter, beam splitter and avalanche photodiode (single photon counting module), respectively.}
\label{Fig1}
\end{figure}

\section{Experimental setup}

Figure \ref{Fig1} shows the schematic diagram of the experiment. The experimental setup is based on a hybrid system of single QDs deposited on an ONF and cooled down inside a cryostat. The QDs used for the experiment were colloidal gradient thick shell type CdSe QDs with an outermost ZnS shell. The QDs were synthesized using a method described in Ref. \cite{Shafi1}. At room temperature, the QDs emit at a wavelength of 640 nm and the emission QE was measured to be 85$\pm$5\% \cite{Shafi1}. ONFs with optical transmission of $>$99\% were fabricated by adiabatically tapering commercial SMFs (SM 600, Fibercore) using a heat and pull technique. ONFs used for the experiment had a waist diameter of 310$\pm$10 nm and a uniform waist length of 2.5 mm. The ONF diameter was chosen to maximize the channeling of the single QD emission into the ONF guided modes with an efficiency as high as 22\% \cite{famsan1,Yalla1}. 

The QDs were deposited on the ONF using a computer-controlled subpico-liter-needle-dispenser system combined with an inverted microscope \cite{Yalla2,Shafi1}. The dispenser consists of a tapered glass-tube containing the QD solution and a tungsten needle with a tip diameter of 5 $\mu$m. Once the needle tip passes through the glass-tube, it carries a small amount of QD solution at its edge and a tiny amount of this solution is deposited on the ONF surface by bringing the tip into contact with the ONF. To achieve reproducible deposition of single QD, the concentration of the QDs was adjusted by diluting the colloidal solution of the QDs in toluene. From various trials, we have estimated an overall success probability of depositing single QD for each trial to be about 60\%.

For the ONF sample presented in this work, ten depositions were performed with a spatial period of 150 $\mu$m. The transmission of the ONF was $>$98\% after the deposition procedure. After the depositions, the ONF/QD hybrid system was installed into a custom-designed optical cryostat \cite{Shafi1}. The ONF transmission was reduced to 90\% during the installation process. The ONF/QD hybrid system was cooled down to 3.7 K via He buffer gas cooling \cite{Shafi1}. The transmission was maintained throughout the cooling process and at the cryogenic temperatures.

The emission characteristics of the QD were measured from the PL photons coupled into the guided mode of the ONF and eventually to a SMF. For measuring the emission characteristics, the QD was excited with a picosecond pulsed-laser with a pulse width of 20 ps FWHM and a wavelength of 532 nm. For photo-charging, the QD was irradiated with a cw laser at a wavelength of 355 nm. The excitation lasers were launched through the optical window of the cryostat and irradiated perpendicular to the ONF. The excitation lasers were focused on the ONF to a spot-size of 19 $\mu$m FWHM using a combination of lenses placed outside the cryostat. The polarization of the excitation lasers was linear with a polarization axis perpendicular to the ONF axis.

The spectral and temporal characteristics of the fiber-guided photons were measured simultaneously at the opposite ends of the fiber. At one end of the fiber, the PL spectra of the QD was measured using an optical multi-channel analyser (OMA) with a resolution of 1.0 meV (0.3 nm). The scattering from the excitation laser coupled into the fiber guided mode was filtered out using a 560 nm long-pass filter (O56, HOYA). At the other end of the fiber, the temporal characteristics were measured using single photon counting modules (APDs, Perkin Elmers) and a time correlated single photon counting system (TCSPC, PicoHarp 300). The single photon characteristics of the emitted photons were inferred from the photon correlations measured using a Hanbury-Brown-Twiss (HBT) setup with two APDs and the TCSPC. As illustrated in Fig. 1, the fiber output was first filtered using a band-pass interference filter (2 nm or 15 nm FWHM) to remove the background photons from the signal, then split into two channels using a beam splitter. The photons in the two channels were measured using two separate APDs and the arrival times of the photons were recorded using a TCSPC. The photon correlations were extracted from the coincidence counts between the two channels for varying delay time. The PL decay profile was measured by observing the temporal correlations between the excitation laser pulse and the PL photons. The repetition rate of the pulsed laser was set to 10 MHz for the measurements of PL spectrum, photon count rate and photon correlations. While, only for the measurement of PL decay profiles, the repetition rate was set to 500 kHz. All data presented here were measured at a temperature of 3.7 K.


\section{Results}

The blue trace in Fig. 2(a) shows the PL spectrum of the single QD measured at 3.7 K with an integration time of 120 s. The spectrum displays well-resolved four peaks labeled as 1, 2, 3 and 4. The observed linewidth (FWHM) of the peak 1 is 3.2 meV and that of the peak 2 is 1.9 meV and they are separated by 16.5 meV. This leads to the assignment of the peaks 1 and 2, as the zero phonon lines (ZPLs) for the exciton state of a neutral CdSe nanocrystal and for the trion state of a charged CdSe nanocrystal, respectively \cite{Shafi1,Louyer,Biadala}. The linewidth of the peak 1 is broader than that of the peak 2, since the upper state of the exciton consists of a bright and a dark exciton state separated by 1.5 meV which was not resolved. The peaks 3 and 4 are red-shifted by 26.5 meV from the peaks 1 and 2, respectively and are attributed to the longitudinal optical phonon (LO-phonon) replica of the exciton and the trion ZPLs \cite{Shafi1,Shafi2,Louyer}. Note that the emission from the exciton and the trion state occurs almost randomly and the occurrence probability reaches a stationary value for the integration time as long as 120 s \cite{Shafi1}. We estimated the trion occurrence probability (TOP) to be 44$\pm3$\% from the ratio between the photon counts emitted by the trion state and the total photon counts from both the trion and the exciton states.

\begin{figure}[t!]
\centering
\resizebox{0.5\textwidth}{!}{%
  \includegraphics{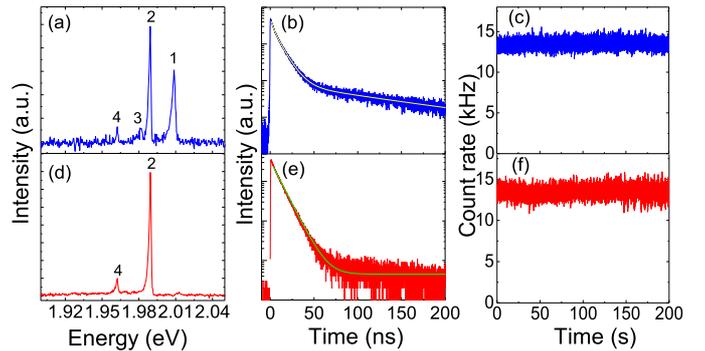}
}
\caption{PL characteristics of a single QD in the ONF/QD hybrid system at 3.7 K. (a) PL spectrum, (b) PL decay profile and (c) photon count rate for a single QD measured before photo-charging. The blue traces are the experimentally measured data. The yellow curve in (b) shows the fit using a sum of three exponential functions. (d) PL spectrum, (e) PL decay profile and (f) photon count rate for a single QD measured after photo-charging. The red traces are the experimentally measured data. The green curve in (e) shows the fit using a single exponential function. The vertical axes in (b) and (e) are in log-scale. All data were measured for an excitation fluence of 6.4 $\mu$J/cm$^2$.
}
\label{Fig2}
\end{figure}

The blue trace in Fig. 2(b) shows the PL decay profile of the single QD at 3.7 K. The decay curve was acquired for an integration time of 200 s with a resolution of 64 ps. The observed decay curve can be well fitted with a sum of three exponential decay functions corresponding to the decay processes for the trion state and the effective fast and slow decay components of the neutral exciton state. The yellow curve in Fig. 2(b) shows the fit. From the fit, we extract three decay times of 2.3$\pm$0.3 ns, 9.6$\pm$0.3 ns and 120$\pm$5 ns corresponding to the effective fast decay component of the neutral exciton state, the decay of the trion state and the effective slow decay component of the neutral exciton state, respectively \cite{Shafi1,Biadala,Labeau}. 

Figure 2(c) shows the photon count rate of the PL of the single QD observed at 3.7 K. The resolution of the time axis is 17 ms. It may be seen that the photon count rate is stable without any intermittency. The observed count rate is 13.0$\pm$0.5 kHz. It should be noted that at room temperature the PL emission shows intermittency due to the lower QE of the trion \cite{Shafi1,Javaux}. The absence of intermittency at cryogenic temperature indicates that both the exciton and trion state have the same QE \cite{Shafi1,Javaux}.
As detailed in Ref. \cite{Shafi1}, the typical overall QE for such QDs is higher than 94\%.


Figure 2(d) shows the PL spectrum of the single QD after the QD was irradiated with 355 nm laser at 5 W/cm$^2$ for about 10 min. The observed spectrum shows only a single emission line corresponding to the trion energy of 1.99 eV (623 nm) along with a LO phonon replica. The absence of the neutral exciton peak suggests that the emission occurs only from the trion state with 100\% TOP. Therefore, it is inferred that the irradiation with 355 nm resulted in permanent photo-charging of the QD.

To further confirm the charged situation, we have measured the PL decay profile of the photo-charged QD. The red trace in Fig. 2(e) displays the PL decay profile of the trion emission. The observed PL decay profile shows a single exponential decay which is consistent with the single upper state model of the trion state \cite{Javaux}. The decay profile is fitted by a single exponential function (green curve) yielding a decay time of 10.0$\pm$0.5 ns. The decay time is one order shorter than the effective slow decay component of the neutral exciton state.

Figure 2(f) displays the typical photon count rate for the photo-charged QD. It may be seen that there is no significant change in the photon count rate compared to Fig. 2(c). This suggests that there is no change in the overall QE of the QD \cite{Shafi1,Fernee1,Javaux}.

The purity of the fiber in-line SPS was measured by carrying out photon correlation measurements of the fiber-guided PL photons from the single QD. To ensure complete relaxation of the excitation between the subsequent laser pulses, the excitation repetition rate was set to be slower than the PL decay rate. The red trace in Fig. 3(a) shows the photon correlations measured for an excitation fluence of 6.4 $\mu$J/cm$^2$ at 10 MHz repetition rate. The coincidences between the outputs of the beam splitter are plotted against the delay time ($\tau$) with a time-bin of 1 ns. It may be seen that periodic peaks with an interval of 100 ns are observed, well corresponding to the repetition rate. However, the central peak at $\tau=0$ is strongly suppressed clearly indicating the single photon characteristics of the source. We estimate the second-order correlation value at $\tau=0$ to be g$^2$(0)= 0.06$\pm0.01$. The value of g$^2$(0) was estimated from the ratio between the area of the peak at $\tau=0$ and the averaged area of side peaks on each side at $\pm100$ ns. The value of g$^2$(0) gives a measure of the purity of the fiber in-line SPS.

\begin{figure}[htp!]
\centering
\resizebox{0.5\textwidth}{!}{%
  \includegraphics{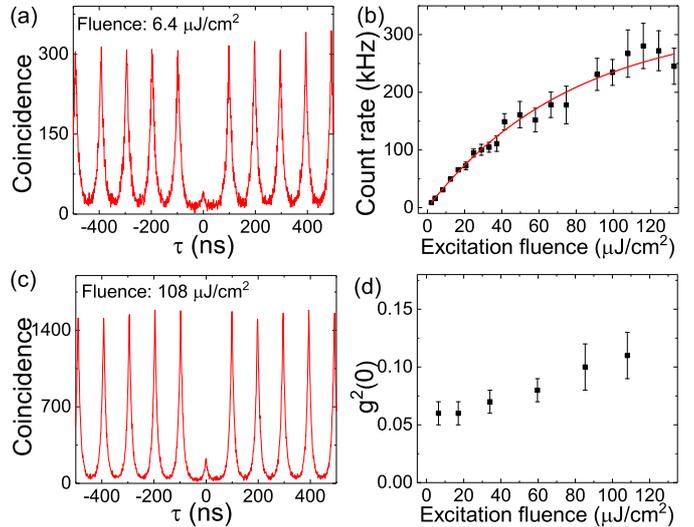}
}
\caption{Photon statistics of the PL of the photo-charged single QD measured at 3.7 K. (a) Photon correlation measurement of the fiber-guided PL photons for an excitation fluence of 6.4 $\mu$J/cm$^2$. (b) Photon count rates (black squares with error bars) at one-end of the fiber measured for different excitation fluences. The red curve is the fit using the saturation model given in Eq. (1). (c) Photon correlation measurement for an excitation fluence of 108 $\mu$J/cm$^2$ which is larger than the ${F}_{sat}$. (d) g$^2$(0) for different excitation fluences.}
\label{Fig3}
\end{figure}

To estimate the brightness (efficiency) of the SPS, the excitation fluence was increased until the photon count rate shows saturation behavior. Figure 3(b) shows the observed photon count rates (black squares with error bars) at one side of the fiber for different excitation fluences. It may be seen that for higher excitation fluence the photon count rate shows a saturation behavior. The observations are fitted using a saturation model for two-level system \cite{Daveau,Waks} given as
\begin{equation}
{I(F)}= {I}_{max}[1-\exp(-\frac{F}{F_{sat}})]
\end{equation}
where ${I}_{max}$ is the saturated photon count rate, ${F}$ and ${F}_{sat}$ are the excitation fluence and the saturation fluence, respectively. The red curve in Fig. 3(b) shows the fit. From the fit, we obtained a maximum count rate of single photons at one side of the fiber to be ${I}_{max}$= $327\pm22$ kHz and a saturation fluence of ${F}_{sat}=79\pm6$ $\mu$J/cm$^2$. 

To confirm the purity of the single photons at the saturation condition, we have measured the photon correlations at higher excitation fluence. Figure 3(c) shows the photon correlations measured at a fluence of 108 $\mu$J/cm$^2$ which is well above the ${F}_{sat}$. It may be seen that the central peak at $\tau=0$ is still strongly suppressed indicating that the single photon characteristics of the SPS is maintained at saturation condition. From the photon correlations, we estimated a g$^2$(0) = $0.11\pm0.02$ at this excitation fluence. Figure 3(d) displays the values of g$^2$(0) estimated for different excitation fluences. It may be seen that g$^2$(0) gradually increases from 0.06 to 0.11 at the saturation condition. This indicates that although the multi-photon emission probability is not significant for the present QD, but at higher excitation fluence it is not negligible. 

We have also recorded the spectral and photon count rate characteristics of the photo-charged single QD at different excitation conditions. Figure 4(a) displays the linewidth (red squares with error bars) and the peak energy (black squares with error bars) of the trion PL spectra measured at different excitation fluences. The linewidths are obtained after correcting for the instrumental response function with a width of 1 meV. It may be seen that there is a systematic increase in the linewidth with increasing excitation fluence. The linewidth increased from 1.1 meV (at 6.4 $\mu$J/cm$^2$) to 3.2 meV (at 108 $\mu$J/cm$^2$). On the other hand, the peak energy shift is negligible (within the resolution limit). Figure 4(b) shows the measured photon count rate at a fluence of 108 $\mu$J/cm$^2$. It may be seen that the fluctuations in the photon count rate are larger compared to that shown in Fig. 2(f). The average photon count rate is estimated to be $280\pm30$ kHz.


\begin{figure}[htp!]
\centering
\resizebox{0.5\textwidth}{!}{%
  \includegraphics{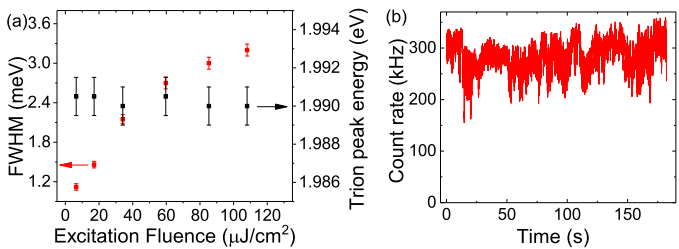}
}
\caption{PL characteristics of the photo-charged single QD at higher excitation fluence. (a) The linewidth (red squares with error bars) and the peak energy (black squares with error bars) of the trion PL spectra measured at different excitation fluences. The linewidths are obtained after correcting for the instrumental response function with an width of 1 meV. (b) Photon count rate measured at an excitation fluence of 108 $\mu$J/cm$^2$ which is larger than the ${F}_{sat}$. }
\label{Fig4}
\end{figure}

\section{Discussion}

The total detection efficiency (${\alpha}$) of the single photons from the ONF to the APD was estimated to be 40\% by taking into account various efficiencies of the optical system which include the ONF transmission (90\%), the filter transmission efficiency (83\%), the fiber-coupling efficiency (90.5\%) to the APD and the detection efficiency of the APD at 620 nm (60\%). By correcting for the transmission and detection efficiencies, we estimate a maximum photon count rate (${\Gamma_{SP}}$ = $\frac{2{I}_{max}}{\alpha}$) of 1.6$\pm$0.2 MHz for the fiber in-line SPS. Here the factor two accounts for the total channeling into both ends of the fiber. The brightness (source efficiency, $\beta$) of the fiber in-line SPS is estimated by relating ${\Gamma_{SP}}$ to the repetition rate ($R$) of the excitation laser as $\beta$ = $\frac{\Gamma_{SP}}{R}$ = 16$\pm$2\%. It should be noted that $\beta$ can be related to the QE ($\eta_q$) of the QD and the channeling efficiency ($\eta$) into the ONF guided modes given as $\beta=\eta_q\eta$. It is reasonable to assume a maximum value for the channeling efficiency ($\eta=22\%$) \cite{famsan1,Yalla1} which would yield a higher $\beta$-value. The reduction in the observed $\beta$ may indicate an overall reduction in the QE of the QD at higher excitation fluence.    

As shown in Fig. 4(b), at high excitation fluence, the photon count rate shows larger fluctuations with a characteristic intermittency behavior. The intermittency may indicate the presence of another emission channel with lower QE. Therefore, the intermittency can effectively reduce the QE of the single QD at high excitation fluence. The intermittency might be related to the emission from Auger-assisted multi-exciton process at high excitation fluence \cite{Javaux,Canneson}. This is also evident from the increase in the $g^2(0)$ value at high excitation fluence. The increase in the linewidth of the PL spectra and onset of the intermittency may also indicate local heating at high excitation fluence \cite{Shafi1,Javaux}. However, the negligible shift in the peak energy suggests that the local temperature was well maintained \cite{Shafi1} and the effect of temperature change may not explain the broadening of the spectral width. The observed increase in the spectral width may be attributed to the enhancement of the spectral diffusion process due to the onset of multi-exciton process \cite{Empedocles}.

Another issue to address at higher excitation fluence is the background induced by the PL of the ONF itself. The level of such background depends on the spot size of the excitation laser and choice of appropriate band-pass filter. In the present cryostat setup, the spot size is limited to 19 $\mu$m and at high excitation fluence, the PL of the ONF induces background photons. However using a 2 nm band-pass filter we can reasonably suppress the background. This is evident from the photon coincidences plotted in Fig. 3(c). We have found that under similar conditions, using a 15 nm band-pass filter will induce background photons which adds a constant offset to the photon coincidences. This offset is $<10$\% of the coincidence peak height. In principle using an objective lens inside the cryostat can reduce the spot-size to $\sim$ 1 $\mu$m to further suppress the background level. It should be noted that it is one of the advantages of using the ONF platform compared to emitters on the bulk substrates where the excitation volume is much larger resulting in significant background from the PL of the substrate itself.

Out of the ten deposited positions on the ONF, we have found single QDs at five positions with similar emission characteristics. The data presented here are representative data measured at one of the positions. For low excitation fluence, we have observed no significant degradation of the emission from the single QDs at least up to 10 days of operation at cryogenic temperatures. However, for excitation fluence higher than the saturation fluence, we have observed degradation of the QD emission within 2 h of continuous operation. To realize a practical device for SPS, further engineering of the QDs will be essential for robust operating conditions.

It should be noted that the QD emission into the ONF guided modes, should depend on the excitation polarization. However, the information about the dipole orientation and azimuthal position of the QD will play a significant role, which is currently not fully explored. At different deposited positions on the ONF, we have observed a variation in the  visibility ranging from 10 to 60\%, measured by changing the excitation polarization from perpendicular to parallel with respect to the ONF axis. The polarization axis was chosen perpendicular to the ONF axis to maximize the photon count rate into the ONF guided modes. 

To enhance the performance of the hybrid ONF/QD system as an SPS, an important step would be to incorporate a cavity structure on the ONF and establish an ONF cavity QED system at cryogenic temperatures. 
Although Purcell enhancement of solid-state quantum emitters on ONF cavities has been successfully demonstrated \cite{Yalla3,Schell}, the experiments were performed at room temperature. As a result, the emission spectral widths were 20-40 nm broad and the cavity enhancement was limited only to a narrow region of the emission spectrum. This limits the overall efficiency of the system as an SPS. In the present hybrid system at cryogenic temperatures, the spectral diffusion width saturates at 3 meV ($\sim$1.0 nm) FWHM, which is comparable to the typical cavity linewidths reported for the ONF cavity QED experiments \cite{Yalla3,Schell,SileFIB,TakeuchiFIB}. As a result, the entire emission band can be strongly enhanced into the ONF cavity mode leading to higher source efficiency. Due to the faster cavity enhanced decay rate, one may also expect a higher photon count rate. Moreover using a one-sided cavity, all photons can be extracted at one end of the fiber. Another key aspect of an SPS is the indistinguishability of the single photons. This will require further narrowing of the spectral width by employing near resonant excitation scheme \cite{Fernee2} and much narrow linewidth ONF cavities as in Ref. \cite{Kali4}.

\section{Conclusions}

In conclusion, we have demonstrated a fiber in-line single photon source based on a hybrid system of colloidal single quantum dots deposited on an optical nanofiber and cooled down to cryogenic temperature (3.7 K). We have shown that the charged single quantum dot exhibits a photo-stable emission of single photons with high quantum efficiency, narrow linewidth (3 meV FWHM) and fast decay time ($10.0\pm0.5$ ns). The single photons were efficiently coupled to the guided modes of the nanofiber and eventually to a single mode optical fiber. The brightness (efficiency) of the single photon source was estimated to be 16$\pm2\%$ with a maximum photon count rate of 1.6$\pm0.2$ MHz and a high single photon purity ($g^2(0)=0.11\pm0.02$). The device can be easily integrated to the fiber networks paving the way for potential applications in quantum networks.

\section{Acknowledgement}
This work was supported by the Japan Science and Technology Agency (JST) through Strategic Innovation Program (Grant No. JPMJSV0918). We acknowledge the contributions of Kazunori Iida and Emi Tsutsumi, in the preparation of the thick shell quantum dot samples.



\end{document}